\documentclass[11pt,a4paper]{amsart}

\usepackage[latin1]{inputenc}
\usepackage[english]{babel}
\usepackage{amsmath}

\usepackage{amssymb, mathabx}
\usepackage[numbers]{natbib}
\usepackage{graphicx}
\usepackage{braket}
\usepackage[pdftex,plainpages=false,colorlinks,hyperindex,bookmarksopen,linkcolor=red,citecolor=blue,urlcolor=blue]{hyperref}

\bibpunct{[}{]}{;}{n}{,}{,}
\usepackage[hmargin=3cm,vmargin={3.5cm,4cm}]{geometry}

\usepackage{color}

\newcommand{\R}{\mathbb{R}}  
\newcommand{\C}{\mathbb{C}} 

\newcommand{\res}{\mathcal{R}es \,}

\begin{document}

\numberwithin{equation}{section}

\allowdisplaybreaks

\title{On infinite series concerning zeros of\\ Bessel functions of the first kind}

    \author{Andrea Giusti$^1$}
		\address{${}^1$ Department of Physics $\&$ Astronomy, University of 	
    	    Bologna and INFN. Via Irnerio 46, Bologna, ITALY.}
		\email{andrea.giusti@bo.infn.it}
	
    \author{Francesco Mainardi$^2$}
    	    \address{${}^2$ Department of Physics $\&$ Astronomy, University of 	
    	    Bologna and INFN. Via Irnerio 46, Bologna, ITALY.}
			\email{francesco.mainardi@bo.infn.it}
			

    \date{\today}

\thanks{Paper published in Eur.~Phys.~J.~Plus (2016) Vol.~131:206, 
\textbf{DOI}: 10.1140/epjp/i2016-16206-4}

\begin{abstract}
A relevant result independently obtained by Rayleigh and Sneddon on an identity on series involving the zeros of Bessel functions of the first kind is derived by an alternative method based on Laplace transforms. Our method leads to a Bernstein function of time, expressed by Dirichlet series, that allows us  to recover the Rayleigh-Sneddon sum. We also consider another method arriving at  the same result based on a relevant formula by Calogero. Moreover, we also provide an electrical example in which this sum results to be extremely useful in order to recover the analytical expression for the response of the system to a certain external input.
\end{abstract} 

\maketitle

\section{Introduction}
	As discussed in our previous paper \cite{AG-FM-2016} dealing with a peculiar mathematical model in linear viscoelasticity,
	 there are strong hints about the possibility of computing the sum of the infinite series 
	\begin{equation}
	S_{\nu} = \sum _{n = 1} ^{\infty} \frac{1}{j_{\nu , \, n} ^{2}} = \frac{1}{4(\nu + 1)} \, , \quad \nu > -1 \, ,  \label{Sneddon} 
	\end{equation} 
	where $j_{\nu , \, n}$ stands for the $n$th positive zero of the Bessel function of the first kind $J_{\nu}$. 
Indeed, in this note the proof is  completely based on the \textit{Laplace transform method}, generalizing our approach followed in  \cite{AG-FM-2016}.   

In Section 2 we present a method 
that, being  inspired  by  our paper  \cite{AG-FM-2016},
is based on  the use of Laplace transform of a ratio of two modified Bessel functions $I_\nu$ of contiguous   order $\nu>-1$.
The inversion   leads to
positive increasing functions  expressed by Dirichlet series
  that  allows us  to recover the Rayleigh-Sneddon result.
  These functions have a complete monotone derivative,
so turn out to be  Bernstein functions.  
       
       In Section 3 we discuss an application of the result within a class of models for electrical ladder networks dual to a class of viscoelastic models discussed in \cite{IC-AG-FM-2016}. In particular, in this section we take advantage of a known analogy between viscoelastic and electrical systems which was widely discussed by Gross and Fuoss in \cite{Gross-Fuoss_1956, Gross_1956-2, Gross_1956-3}.
       
       In Section 4 we briefly discuss an alternative method 
       based on a relevant formula due to 
       Calogero \cite{Calogero LNC77} of which we became aware 
       only in the later stage of this work. 
  
    Concluding remarks are found in Section 5 where we outline the different motivation of the two approaches which appear to be somehow correlated. Moreover, in this section we also give some hints for potential future research concerning the analogous electrical model discussed in Section 3.
    
  Detailed mathematical aspects of our approach discussed in Section 2 are presented in two Appendices for reader's convenience.

\section{The Rayleigh-Sneddon result based on Laplace transform} 

Let us consider the following Laplace transform defined in the complex 
$s$-plane,
	\begin{equation} \label{eq_LT_rel}
	\widetilde{F} _{\nu} (s) = \frac{2 (\nu + 1)}{s\sqrt{s}} \frac{I_{\nu + 1} (\sqrt{s})}{I_{\nu} (\sqrt{s})} \, , \quad \nu > - 1 \, . 
	\end{equation}
	In view  of the power series representation for the Modified Bessel functions of the first kind, 
see e.g. \cite{AS 1965,TEMME 1996}:
\begin{equation}\label{BESSEL series}
I_{\nu} (z) = \left( \frac{z}{2} \right) ^{\nu} \, \sum _{k=0} ^{\infty} \frac{1}{k! \, \Gamma (\nu + k + 1)} \left( \frac{z}{2} \right) ^{2 k} 
\,  \end{equation} 
we recognize that  $\widetilde{F} _{\nu} (s)$
turns out to be a single-valued function,
with a simple pole  in $s=0$  with residue 1, and infinite  poles in the points where 
\begin{equation}
 I_{\nu} (\sqrt{s}) = 0 \quad \Longleftrightarrow \quad J_{\nu} (\lambda) = 0 \, ,
 \end{equation}
 having renamed  $\sqrt{s} = - i \lambda$ with $\lambda  >0$
 and denoted by $J_\nu$ the standard Bessel functions of the first kind. 
 For $\nu > -1$ the zeros
 of $J_\nu$ are known to be all simple and real so that  the  poles of 
 $\widetilde{F} _{\nu} (s)$ (different from zero) are   
located on the negative real axis. They are given by
\begin{equation}\label{poles}
s_{\nu ,n} = - j_{\nu , \, n} ^{2}\,, \quad  n = 1, 2, \ldots
\end{equation}
where  $j_{\nu , \, n}$ 
stands for the $n$th {\it positive zero} of  $J_\nu$.
Of course, for  $\widetilde{F} _{\nu} (s)$ 
 the infinity is a singular point being a limit  point of poles 
but for $Re \{s\} \to +\infty$ the function goes to zero being   
\begin{equation} \label{LIMIT-infty}
\widetilde{F} _{\nu} (s) \sim 2(\nu+1)\,  s^{-3/2}\,, \quad 
Re \{s\} \to+\infty,
\end{equation}
as we recognize from the known asymptotic representation of the modified Bessel function 
\begin{equation}
I_\nu(z) \sim \frac{e^z}{\sqrt{2\pi}\, z^{1/2}}\,, 
\quad  |z| \to \infty, \; |\hbox{arg} \,z| <\frac{\pi}{2}\,.
\end{equation}    
Inverting 	$\widetilde{F} _{\nu} (s)$ by using the residue theorem,
see for details Appendix A,  we obtain
the  original function  $F _{\nu} (t)$ (in the time domain)  as 
\begin{equation}\label{Dirichlet_1}
F _{\nu} (t) = 
1- 
4 (\nu +1)\,
 \sum_{n=1}^\infty 
\frac{\exp \left( - j^2_{\nu , \, n} \, t \right)}{j^2_{\nu , \, n}}
  \, , \quad t > 0\,.
 \end{equation}
The series of exponentials in the R.H.S is a particular type of
 Dirichlet series of which we prove the absolute convergence for $t>0$ in Appendix B.
      
	From this representation by means of  a  Dirichlet series we recognize that 
$F_\nu(t)$ is a locally integrable, positive and increasing   function   for $t>0$,   with infinite a derivatives alternating in sign.
The local integrability is easily proved by recognizing   that
  \begin{equation} \label{asymptotic}
  F_\nu(t) \sim 4(\nu+1) \,\frac{ t^{1/2}}{\sqrt{\pi}}\,, \quad  t \to 0^+\,,
  \end{equation} 
in virtue of  the asymptotic behaviour of  the Laplace 
 transform $\widetilde F_\nu(s) $ outlined in 
 Eq. (\ref{LIMIT-infty}).
  
  Using a precise mathematical terminology,  the  function  $F_\nu(t)$ is  classified as a Bernstein function  and its derivative 
$F^\prime_\nu(t)$ as a completely monotone (CM) function. 
For more details on these  classes of functions
 we may refer the interested reader to the excellent  monograph by Schilling et al.  \cite{ Schilling-et-al 2012}.

Now setting to zero the limiting value $F_\nu(0^+)$ in
Eq. (\ref{Dirichlet_1})
we promptly derive the result by Rayleigh and Sneddon expressed in Eq. (\ref{Sneddon}).


\section{An electrical application based on electro-mechanical analogies}
	In 1956, B. Gross and R. M. Fuoss first presented a formal correspondence between the analytical description of electrical ladder structures and viscoelastic systems. In particular, in \cite{Gross-Fuoss_1956} they considered a model for the deformation of an elastic rod in a viscous medium.  To be more specific, let $\sigma = \sigma (t, x)$ and $\varepsilon = \varepsilon (t, x)$ be the stress and the strain functions at a point $x$ and a time $t$, respectively. Then, the viscoelastic model in \cite{Gross-Fuoss_1956} is given by the following stress-strain relations:
	\begin{equation} \label{eq-a}
	\sigma = e \, \frac{\partial \varepsilon}{\partial x}
	\, , \qquad
	\frac{\partial \sigma}{\partial x} = \eta \, \frac{\partial \varepsilon}{\partial t}
	\end{equation}
	where $1/e$ and $\eta$ are respectively the spring modulus and the viscosity constant.
	
		Clearly, a priori such a model would not be linear, but it can be linearized (as above) by considering the case of small deformations. As argued in \cite{Gross-Fuoss_1956}, truly linear systems can be given in terms electrical analogs of mechanical models. Indeed, an example of electrical analog to the viscoelastic system discussed by Gross and Fuoss is provided by a finite cable in which a uniform wire of specific resistance $\rho$ is enclosed into a dielectric material of fixed thickness.  For such an electrical system we have that:
		\begin{equation} \label{eq-b}
	I = \frac{1}{\rho} \, \frac{\partial V}{\partial x}
	\, , \qquad
	\frac{\partial I}{\partial x} = c \, \frac{\partial V}{\partial t}
	\end{equation}
	where $I = I(t,x)$ and $V = V(t,x)$ are respectively the current and the potential. Moreover, $\rho$ and $c$ are respectively the resistance and the capacitance per unit of length. 
	
	Now, comparing Eq. (\ref{eq-a}) and Eq. (\ref{eq-b}) we can immediately read off the prescriptions commonly used to implement electro-mechanical analogies. Indeed, as discussed in \cite{Gross-Fuoss_1956, Gross_1956-2, Gross_1956-3}, we heve
	\begin{equation*}
	\left.
	\begin{aligned}
	& \sigma \,\, \mbox{stress}\\
	& \varepsilon \,\, \mbox{strain}\\
	& E \,\, \mbox{elastic modulus}\\
	& \eta \,\, \mbox{viscosity}\\
	\end{aligned}
	\right\} 
	\quad
	\Longleftrightarrow
	\quad
	\left\{
	\begin{aligned}
	& I \,\, \mbox{current}\\
	& V \,\, \mbox{potential}\\
	& 1/R \,\, \mbox{conductance}\\
	& C \,\, \mbox{capacitance}\\
	\end{aligned}
	\right.
	\end{equation*}
	
	Now, if we start off with a more general viscoelastic system, its corresponding analogous electrical system would not be described simply in terms of single electrical component. Indeed, in general we have that a viscoelastic system formally corresponds to a class of electrical ladder networks given by the following formal duality
	 \begin{equation*}
	\left.
	\begin{aligned}
	& \mbox{Spring}\\
	& \mbox{Dashpot}\\
	\end{aligned}
	\right\} 
	\quad
	\Longleftrightarrow
	\quad
	\left\{
	\begin{aligned}
	& \mbox{Resistor}\\
	& \mbox{Capacitor}\\
	\end{aligned}
	\right.
	\end{equation*}
	
	Let us consider the function $\widetilde F _{\nu} (s)$ in Eq. (\ref{eq_LT_rel}). As discussed in \cite{IC-AG-FM-2016}, the function
	\begin{equation}\label{eq-c}
	\widetilde \Phi _\nu (s) = s \, \widetilde F _{\nu} (s) = \frac{2 (\nu + 1)}{\sqrt{s}} \frac{I_{\nu + 1} (\sqrt{s})}{I_{\nu} (\sqrt{s})} \, , \quad \nu > - 1 \,
	\end{equation}
	represents the Laplace transform of the relaxation memory function 	$\Phi_\nu(t)$ for a general class of viscoelastic models.
 Then in time domain we get by using Eq.(\ref{Dirichlet_1}) 
	\begin{equation}\label{eq-cbis}
	\Phi_\nu(t) = \frac{dF_\nu}{dt}(t) =
	4 (\nu + 1) \, \sum _{n=1} ^\infty \exp \left( - j_{\nu, n} ^2 t \right)\,.
	\end{equation}
Following the general description of linear viscoelasticity displayed in \cite{Mainardi BOOK2010} 
for bodies  in rest for $t<0$, 
one immediately gets the stress-strain relation in the Laplace domain for the class of models corresponding to the relaxation memory function in Eq. (\ref{eq-c}), i.e.
	\begin{equation}
	\widetilde \sigma (s) = 
	s \, \widetilde G _\nu (s) \, \widetilde \varepsilon (s) =	
	(1 - \widetilde \Phi _\nu (s)) \, \widetilde \varepsilon (s)
	\end{equation}
	where $\widetilde G _\nu (s)$ is the Laplace transform of the so called relaxation modulus $G_\nu(t)$ with a finite limiting  value at $t=0$.  
	We note that, for the sake of convenience, we consider all quantities in   non-dimensional form   after a suitable scaling so that we agree to take  
	$G_\nu(0^+)=1 $ .
	Inverting  $\widetilde G _\nu (s)$ we get
	\begin{equation}
	G _\nu (t) = 1 - \int _0 ^t \Phi _\nu (\tau) \, d \tau = 
 1 - 4 (\nu + 1) \, \sum _{n=1} ^\infty \frac{1 - \exp \left( - j_{\nu, n} ^2 t \right)}{j_{\nu, n} ^2} \, .
	\end{equation}	
Using Eq. (\ref{Sneddon}) this expression 
immediately turns into
	\begin{equation}
	G_\nu (t) = 4 (\nu + 1) \, \sum _{n=1} ^\infty \frac{\exp \left( - j_{\nu, n} ^2 t \right)}{j_{\nu, n} ^2} \, .
	\end{equation}
	Alternatively we can get $G_\nu(t)$ from Eq. (\ref{Dirichlet_1}) noting 
\begin{equation}
	G_\nu(t) = 1 - F_\nu(t)\,.
	\end{equation}
	
	Now, the stress-strain relation in the time domain and in non-dimensional form is given by
	\begin{equation}
	\sigma (t) = \varepsilon (t) + ({\dot G} _\nu \ast \varepsilon)(t) \, ,
	\end{equation}		
	where $\ast$ represents the convolution product\footnote{Let $f(t)$, $g(t)$ be causal functions (i.e. $f(t) \equiv f(t) \, \Theta (t)$, where $\Theta (t)$ is the Heaviside distribution), as in case of our concern, we have that
$$ (f \ast g) (t) = \int _\R f(t - \tau) \, g(\tau) \, d\tau = \int _0 ^t f(t - \tau) \, g(\tau) \, d\tau $$} and the dot represents the derivative with respect to the argument of the function. If we then consider the electro-mechanical analogy discussed above we get
	\begin{equation}\label{eq-d}
	I (t) = V (t) + ({\dot G} _\nu \ast V)(t) \, ,
	\end{equation}
	which represents a characteristic equation for some peculiar electrical ladder networks dual to the viscoelastic model introduced above. 
	From the mathematical and physical points of view, it is interesting 
to  study the  response  of the current 	$I(t)$ to a step potential, i.e. $V (t) = \Theta (t)$. 
	 Therefore, plugging the input potential into Eq. (\ref{eq-d}) and integrating by parts we get
\begin{equation}
I (t) = G_{\nu} (t) = 4 (\nu + 1) \, \sum _{n=1} ^\infty \frac{\exp \left( - j_{\nu, n} ^2 t \right)}{j_{\nu, n} ^2} \, ,
\end{equation}
which clearly describes a relaxation process featured by a completely monotone behavior.

The  response of the current is thus decaying exponentially for large times.
However it is surprising to note that the initial decay rate of the current
expressed by its time derivative 
	\begin{equation}
	\frac{dI}{dt} =   \frac{dG_\nu}{dt} = - \Phi_\nu(t)
	\end{equation} 
 is very strong because from  Eq. (\ref{asymptotic})
 we get    	
	\begin{equation}
	\Phi _\nu(t) \sim 2(\nu+1) \,\frac{ t^{-1/2}}{\sqrt{\pi}}\,, 
	\quad  t \to 0^+\,.
	\end{equation}	 
	This asymptotic behaviour at $t=0^+$ is peculiar 
	of  our electro-mechanical system; it is essentially due to the  infinite (but discrete) distribution of simple relaxation elements
	of  the Dirichlet series. 
	In other words our  Dirichlet series generalizes the finite Prony series commonly adopted in   the analysis of relaxing electrical and mechanical systems.      


\section{The Rayleigh-Sneddon sum based on Calogero's formula}

We note that Eq.(\ref{Sneddon}) is a particular case of  more general 
formulas due to Lord Rayleigh in 1874, reported in the classical
treatise by Watson \cite{Watson BOOK1944}, page  502, concerning the zeros of the Bessel functions.
 Related formulas were  proved in 1960 by Sneddon \cite{Sneddon} by a different approach, including again our  Eq. (\ref{Sneddon}). 

Recently we became aware of the paper by Baricz et al \cite{Baricz-et-al JMAA15} in which the authors have pointed out  a  formula derived in 1977 by Calogero 
\cite{Calogero LNC77} from which we can derive our 
particular case of the general results by Rayleigh and Sneddon   after a few passages.
Indeed, Calogero,  based  on the well known infinite product representation of the Bessel functions of  the first kind 
\begin{equation}
     J_\nu (x) = \frac{(x/2)^\nu}{\Gamma(\nu+1)}\,
     \prod_{n=1}^{\infty} \left( 1 - \frac{x^2}{j_{\nu,n}^2} \right)\,,
     \quad \nu>-1\,,
\end{equation}      
 and by means of  an equivalent form of the Mittag-Leffler expansion, proved that
 \begin{equation}
 \frac{  J_{\nu+1} (x)}{   J_\nu (x) } =
 \sum_{n=1}^{\infty}  \frac{2x}{{j_{\nu,n}^2} -x^2}\,,
 \quad \nu>-1\,.
 \end{equation}
Then, let us  take the following limit as $x \to 0$
\begin{equation}\label{CALOGERO eq_lim1}
\lim_{x\to 0} 
 \frac{1}{2x}\, \frac{  J_{\nu+1} (x)}{  J_\nu (x) } =
\sum_{n=1}^{\infty}  \frac{1}{{j_{\nu,n}^2} }= S_\nu\,,
 \quad \nu > -1\,.
\end{equation}
Now,  recalling  the Taylor expansion of the Bessel functions of the first kind,
\begin{equation}
J_{\nu} (z) = \left( \frac{z}{2} \right) ^{\nu} \, 
\sum _{k=0} ^{\infty} \frac{(-1)^k}{k! \, 
\Gamma (\nu + k + 1)} \left( \frac{z}{2} \right) ^{2 k} \, ,
\end{equation}
the limit in Eq. (\ref{CALOGERO eq_lim1}) turns out to be after simple calculations 
\begin{equation}\label{CALOGERO eq_lim2}
\lim_{x\to 0} 
 \frac{1}{2x}\, \frac{  J_{\nu+1} (x)}{  J_\nu (x) } =
\frac {1}{4(\nu+1)}\,,
 \quad \nu > -1\,.
\end{equation}
Comparing the two limits in Eqs.  (\ref{CALOGERO eq_lim1}),
(\ref{CALOGERO eq_lim2}) we obtain the Rayleigh-Sneddon sum.	
	
\section{Concluding remarks}

It is straightforward to note that the derivation of the Rayleigh-Sneddon result 
discussed in Section 4  is  more  direct than ours   based on the Laplace transforms, presented in Section 2.
By the way   our method allows to define  locally integrable functions of time, $F_\nu(t)$,   
expressed as Dirichlet series  related to the positive zeros of the  Bessel functions of the first kind  of order $\nu>-1$.
We recognize that these  functions  are of    Bernstein type, 
  and thus suitable to characterize  creep processes in linear viscoelasticity,  see e.g. \cite{Mainardi BOOK2010}.
	As a consequence, their derivatives (with minus sign) are thus completely monotonic and hence suitable for relaxation processes in electro-mechanical systems as outlined along this paper. 
  
   The two methods, even though they appear to correlated,
   arise from different motivations.
   
   On the one hand, the approach discussed in Section 4 is essentially  based on a noteworthy formula by Calogero that was motivated by the researches about the connection
between the motion of poles and zeros of special solutions of partial differential equations and many-body problems as outlined in 
\cite{Baricz-et-al JMAA15}.

  On the other hand,  our approach discussed in Section 2 was inspired by our analysis  
   dealing with a peculiar mathematical model in linear viscoelasticity,
   see \cite{AG-FM-2016}.  
 This model was characterized in the Laplace domain by ratios of modified Bessel  of contiguous order 0,1  and 1,2,  leading to completely monotone and Bernstein functions.
 In  \cite{AG-FM-2016}  these functions were  expressed in terms of  Dirichlet series involving the positive zeros of the Bessel functions  $J_0$ and $J_2$.
  Extending  to generic orders $\nu>-1$ but always using the Laplace transforms,  in the present note we  are  able to prove the  mathematical result by Rayleigh and Sneddon but also, in a forthcoming work \cite{IC-AG-FM-2016}, to introduce a  new class of viscoelastic models. 
  
	Finally, in Section 3 we have been able to provide a general class of electrical ladder networks featuring the properties of being dual to a general class of viscoelastic media described in \cite{IC-AG-FM-2016}. Then, it would be interesting for future research to investigate the dynamical evolution equations in space and time arising from these kinds of electro-mechanical models.
  
\vskip 0.5 cm  
  
\noindent \textbf{Acknowledgements.} 
	The authors would like to thank Ivano Colombaro for the many discussions on the preliminary draft of this paper. We are also grateful to Arpad Baricz for pointing out some interesting historical facts connected with Eq. (\ref{Sneddon}) and to the anonymous reviewer for the thorough reading and constructive comments on our manuscript. The work of A.G. and F. M. has been carried out in the framework of the activities of the National Group for Mathematical Physics (GNFM, INdAM).

 \section*{Appendix A}
	Here we show  the details how to obtain $F _{\nu} (t)$
in Eq. (\ref{Dirichlet_1}) 	by inverting the  Laplace transform 
$\widetilde  F_{\nu} (s)$ in Eq. (\ref{eq_LT_rel}).
We need to 	 evaluate  the Bromwich Integral:
\begin{equation}
F _{\nu} (t) = \frac{1}{2 \pi i} \int _{Br}  \widetilde{F} _{\nu} (s) \, e^{st} \, ds \, , \tag{A1}
\end{equation}
by applying the Residue Theorem.
\\
\begin{proof}
Let us first note that $s=0$ is a simple pole for $\widetilde F_\nu(s)$
whose residue is 1  as it is  deduced  by  computing the limit of
 $s\,\widetilde F_\nu(s)$ as $s\to 0 $. In fact this result is easily obtained by taking  the first term of the Taylor series
 (\ref{BESSEL series}) 
 for $I_\nu$ and $I_{\nu+1}$ around $s=0$.
    
Let us then  recall that  the simple poles 
$s_{\nu,n}$
exhibited   by  $\widetilde{F} _{\nu} (s)$ on the negative real axis are
 the zeros of $I_\nu(\sqrt{s})$ and hence  
 related to the positive zeros $j_{\nu , \, n}$ 
of the Bessel function of the first kind 
$J_\nu(z)$   via Eq. (\ref{poles}), that we repeat hereafter for convenience:
\begin{equation}
s_{\nu,n}  = -j_{\nu , \, n}^2\,,  \quad n=1,2, \ldots. 
\tag{A.2}
\end{equation}
We thus  have
\begin{equation}
F _{\nu} (t) =  1+
\sum _{s_{\nu, n}}
 \res \left\{ \widetilde{F} _{\nu} (s) \, e^{st} 
 \right\}  _{s=s _{\nu,n}} 
 = 1+ \sum _{n=1} ^\infty \res \left\{ \frac{2 (\nu + 1)}{s\sqrt{s}} \frac{I_{\nu+1} (\sqrt{s})}{I_{\nu} (\sqrt{s})} \, e^{st} 
 \right\} _{s=s _{\nu, n}} , \tag{A3}
\end{equation}
where the $n$-th  residue can be computed as follows: 
\begin{equation}
\begin{split}
& \res \left\{ \frac{2(\nu+1)}{s\,\sqrt{s}} \frac{I_{\nu + 1} (\sqrt{s}) }{I_{\nu} (\sqrt{s})} \, e^{st}\right\} _{s = s_{\nu, n}} 
= \lim_{s \to s_{\nu,n}}  (s - s_{\nu, n}) \,
 \frac{2 (\nu + 1)}{s\,\sqrt{s}} 
\frac{I_{\nu + 1} (\sqrt{s}) }{I_{\nu} (\sqrt{s})} \,e^{st} \\
&=  \lim_{s \to s_{\nu,n}} \, \frac{2 (\nu + 1)}{s\,\sqrt{s}} 
\frac{I_{\nu + 1} (\sqrt{s})}
{I^\prime_{\nu} (\sqrt{s})/(2\sqrt{s})}\, e^{st}=  
4 (\nu +1) \, \frac{\exp \left( s_{\nu, n} t \right)}{ s_{\nu, n}} \, . 
\end{split} \tag{A4}
\end{equation}
Above  we have used the property that if $s_{\nu,n}$ is a zero of the modified Bessel function $ I_\nu$,  then we have,
\begin{equation}
I_\nu^\prime(s_{\nu, n}) = I_{\nu+1} (s_{\nu,n}) \,,
\tag{A5}
\end{equation}
as we deduce from the general identity for modified Bessel functions, see e.g.  \cite{AS 1965}, 
$$ I'_\nu(z) = \frac{\nu}{z} I_\nu(z) + I_{\nu+1} (z)\,.$$
Thus, we finally get
\begin{equation}
F _{\nu} (t) = 1+ \sum _{n=1} ^\infty 
\res \left\{ \widetilde{F} _{\nu} (s) \, e^{st} \, ; \,
 s = - j_{\nu , \, n} ^{2} \right\} = 
1-
 4 (\nu + 1) \, \sum_{n=1} ^\infty 
 \frac{\exp \left( - j_{\nu , \, n} ^{2} \, t \right)}
{ j_{\nu , \, n} ^{2}}
  \, . \tag{A6}
\end{equation}
in agreement with Eq. (\ref{Dirichlet_1}).
\end{proof}

\section*{Appendix B}
Here we show 
that the Dirichlet series (\ref{Dirichlet_1}) is absolutely convergent for $ t> 0$ basing the proof on a statement of the  classical treatise by Hardy and Riesz  \cite{Hardy-Riesz 1915}.

\begin{proof}
Consider a  Dirichlet series in the $z$-complex plane:
\begin{equation} \label{Dirichlet_2}
f(z) = \sum _{n=1} ^\infty a_n \, \exp \left(- \alpha _n z \right) \,\, , \qquad \quad z \in \C \, . \tag{B1}
\end{equation}
We have 
convergence  and absolute convergence in the right half planes
$Re\{z\} > \sigma _c $ and $Re\{z\}> \sigma _a$, respectively,
with $ \sigma _a\ge  \sigma _c$. 
The abscissa of convergence  $\sigma _c$
and the abscissa of absolute convergence 
$ \sigma _a$ satisfy the following condition:
\begin{equation}
0 \leq \sigma _a - \sigma _c \leq d = \limsup _{n \to \infty} \frac{\ln n}{\alpha _n} \, . \tag{B2}
\end{equation}
If $d=0$, then
\begin{equation}
\sigma \equiv \sigma _c = \sigma _a = \limsup _{n \to \infty} \frac{\ln \left| a_n \right|}{\alpha _n} \, . \tag{B3}
\end{equation}
In the case of our concern, $a_n = 1/j_{\nu , \, n} ^{2}$ and 
$\alpha _n = j_{\nu , \, n} ^{2}$. 
Then, we have to understand the behaviour of the coefficients 
$j_{\nu, \, n}$ for $n \gg 1$.
\\
Considering the asymptotic representation:
\begin{equation}
J_{\nu} (x) \overset{x \gg 1}{\sim} \sqrt{\frac{2}{\pi x}} \, \cos \left( x - \frac{(2 \nu + 1) \pi}{4} \right) + o \left(x^{- 3/2}\right ) \, ,
 \tag{B4}
\end{equation}
we can eventually get to the following conclusion:
\begin{equation}
J_{\nu} (j_{\nu , \, n}) = 0 \, , \,\,\, \mbox{for} \,\, n \gg 1 \quad \Longrightarrow \quad j_{\nu , \, n} \, \propto \, n \, , \,\,\, \mbox{for} \,\, n \gg 1 \, . \tag{B5}
\end{equation}
Thus,
\begin{equation}
\frac{\ln n}{\alpha _n} = \frac{\ln n}{j_{\nu , \, n} ^{2}} \,\, \overset{n \gg 1}{\sim} \,\, \frac{\ln n}{n^2} \,\, \overset{n \to \infty}{\longrightarrow} \,\, 0 \, , \tag{B6}
\end{equation}
from which we deduce that $d=0$.\\
Finally,
\begin{equation}
\sigma \equiv \sigma _c = \sigma _a = \limsup _{n \to \infty} \frac{\ln \left| a_n \right|}{\alpha _n} = 0 \, . \tag{B7}
\end{equation}
This result allows us to conclude that the Dirichlet series in Eq. (\ref{Dirichlet_1}) is absolutely convergent for $t > 0$.
\end{proof}

\end{document}